\documentclass[referee,english]{raa}
\usepackage{graphicx,times,epstopdf,color,caption}
\usepackage{natbib}
\usepackage{amssymb,amsmath}
\usepackage{babel}
\usepackage{booktabs}
\usepackage{threeparttable}
\usepackage{hyperref}

\begin{document}

   \title{Gaia Calibrated UV Luminous Stars in LAMOST
}

 \volnopage{ {\bf 20XX} Vol.\ {\bf X} No. {\bf XX}, 000--000}
   \setcounter{page}{1}

   \author{Yu Bai\inst{1}, JiFeng Liu\inst{1,2}, Song Wang\inst{1}
   }

   \institute{ Key Laboratory of Optical Astronomy, National Astronomical Observatories, Chinese Academy of Sciences,
       20A Datun Road, Chaoyang Distict, Beijing 100012, China; \it{ybai@nao.cas.cn}\\
        \and
             College of Astronomy and Space Sciences, University of Chinese Academy of Sciences, Beijing 100049, China
\vs \no
   {\small Received 20XX Month Day; accepted 20XX Month Day}
}

\abstract{
We take advantage of the Gaia data release 2 to present 275 and 1,774 ultraviolet luminous stars in the FUV and the NUV.
These stars are 5$\sigma$ exceeding the centers of the reference frame that is built with over one million
UV stars in the log $g$ vs $T_\textrm{eff}$ diagram.
The Galactic extinction is corrected with the 3D dusty map.
In order to limit the Lutz-Kelker effect to an insignificant level, we select the stars with the relative uncertainties of
the luminosity less than 40\% and the trigonometric parallaxes less than 20\%.
We cross-identified our sample with the catalogs of RR Lyr stars and possible white dwarf main-sequence binaries, and find they
compose $\sim$ 62\% and $\sim$ 16\% of our sample in the FUV and NUV, respectively.
This catalog provides a unique sample to study stellar activity,
spectrally unresolved compact main-sequence binaries and variable stars.
}

   \authorrunning{Y. Bai et al. }            
   \titlerunning{Gaia Calibrated UV Luminous Stars in LAMOST}  
   \maketitle

%
\section{Introduction}           
\label{sec:int}
The majority of the stars in our Galaxy is cool, emitting much of the stellar electromagnetic radiation in the
visible or near-infrared part of the spectrum. These cool stars are defined by the effective temperatures that
are the thermal temperatures estimated from stellar photospheres. The higher regions of the stellar atmosphere
(chromosphere, transition region, and corona) are often dominated by more violent non-thermal physical processes mainly
powered by the magnetic field \citep{Bai18}. These processes could produce high energy emissions in the UV or X-ray
part of the spectrum, which are known as stellar activity.

The stellar activity could release a total energy higher than 10$^{34}$ ergs, which leads to discrepancies
between observations and the stellar theocratical models. Such discrepancy is severe in the UV band that is particularly sensitive
to hot plasma emission ($\sim$ 10$^4-$10$^6$ K). Therefore, the UV domain is ideal
for investigating stellar activity, and its availability has been explored for Sun-like stars \citep{Findeisen11}
and M dwarfs \citep{Jones16}. An observational reference frame in the UV is essential to characterize the
stellar activity and further to identify the truly peculiar stars, which is still poorly understood due to the previous
small size of samples.

The UV luminous stars are outliers that don't fit such UV reference frame. They are probably flaring stars due to
magnetic reconnection. The stellar magnetic field is though to be generated and maintained by a stellar dynamo \citep{Wright11},
or interaction in a binary star \citep{Simon80} or in a star-planets system \citep{Ip04}. They could also be the stars
with very hot atmospheres, e.g., accreting pre-main sequence stars \citep{Eaton95}, hot subdwarfs \citep{Wang17} and variable stars \citep{Sesar10,Bai18}.
The excessive UV emission may be originated from the spectrally unresolved companions that are active late-type
stars \citep{Yang17}.

On the other hand, the main-sequence stars around compact objects that cannot be resolved by the LAMOST spectra,
could also emit excessive UV photons \citep{Jao14}.
Such UV emission could be originated from the accretion disks round white dwarfs \citep{Gansicke97,Gansicke01},
neutron stars or black holes. The ratios between X-ray and UV luminosity are about 0.1-100 for quiescent accretion disks around
neutron stars and black holes \citep{Hynes12,Cackett13,Froning14}. This UV emission can provide important
information to study compact objects in binary systems.
Therefore, these UV outliers give us an unique sample that enables us to further investigate the stellar activity
and the evolution of stars and binaries.

\citet{Bai18} presented a UV catalog of over three million stars selected from the data release 3 of
the LAMOST survey \citep{Cui12}, in which about two third were detected by the $Galaxy$ $Evolution$ $Explorer$ ($GALEX$;
\citealt{Morrissey07}). We take advantage of the catalog to study the UV reference frame with the help of the Gaia
data release two (DR2; \citealt{Gaia16,Gaia18a}). We present the data for the calculation of UV luminosity in Section \ref{sec:dat}.
The selection of the UV luminous stars is presented in Section \ref{sec:res}. Section \ref{sec:sum} gives a summary.

\section{Data}
\label{sec:dat}
We calculate the average visit magnitudes in \citet{Bai18} and extract the stellar parameters of the effective temperature ($T_\textrm{eff}$)
and the surface gravity (log $g$). Gaia DR2 provides parallaxes that can be used to obtain distance information.
The UV stars are cross-matched to the Gaia DR2 with a match radius of {2\arcsec}.
Estimating distance directly from the trigonometric parallax may suffer the Lutz$-$Kelker Effect (LKE), which was discussed in detail by \citet{Trumpler53},
and then parametrically formalized by \citet{Lutz73}. The effect is defined as an offset between the average absolute magnitudes
for classes of stars as determined from trigonometric parallax samples and the true mean absolute magnitude for that stellar
class \citep{Lutz73,Sandage16}. The bias is found to be significant for stars with relative high parallax uncertainty, $\sigma_{\pi}/\pi \gtrsim$ 20\%.
However, the universal application of the correction for the bias has been challenged in more recent years \citep{Smith03,van07,Francis14}.
In order to limit the LKE to an insignificant level, we select the stars with relatively small uncertainties of the trigonometric parallaxes.

\begin{figure}
   \centering
    \includegraphics[width=.8\textwidth]{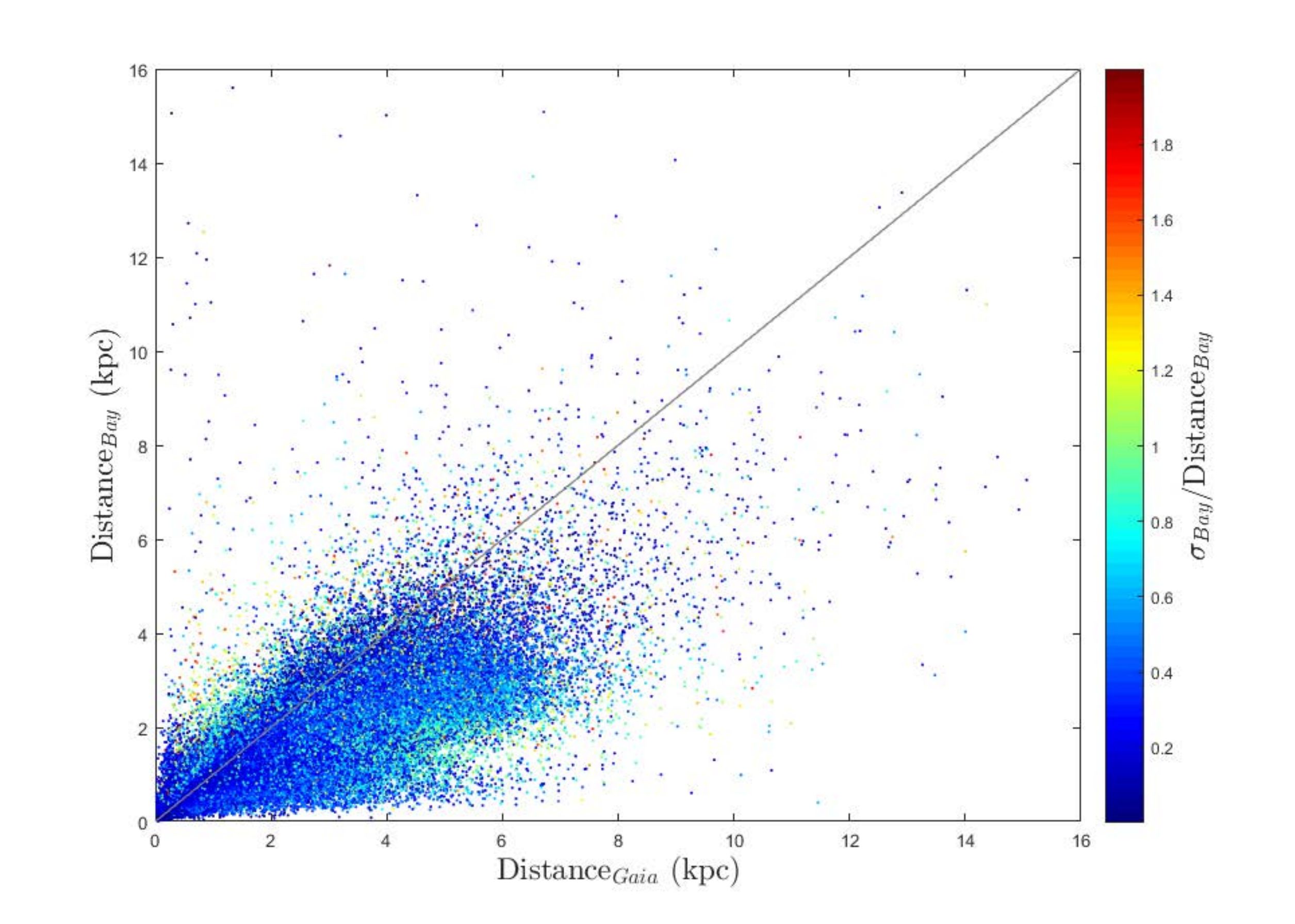}
  \caption{Comparison of distances estimated with the Bayesian method with those from Gaia. The color bar stands for the relative uncertainties of the
           Bayesian distances. }
  \label{Bay} 
\end{figure}

The Bayesian method developed by \citet{Burnett10} and \citet{Binney14} has been used for stars in the RAVE survey, which has demonstrated
the ability to obtain accurate distance and extinction. \citet{Wang16} measured extinctions and distances using this method for stars with valid stellar
parameters in the second data release of the LAMOST survey. They used the spectroscopic parameters $T_\textrm{eff}$, [Fe/H] and log $g$, and 2MASS
photometry to compute the posterior probability with the Bayesian method. We compare their distances to those from Gaia in Fig. \ref{Bay}.
Here, we plot the stars with relative uncertainties less than 20\% for the Gaia distances, and don't constrain the relative uncertainties of
the Bayesian distances \citep{Bai18}. The distances from the Bayesian method are underestimated probably due to the uncertainties in the stellar parameters,
if they are not well constrained in the LAMOST pipeline \citep{Wu11}.
The trigonometric parallaxes in the Gaia DR2, however, have good consistency with those from other methods \citep{Gaia18b}.

\begin{figure}
   \centering
    \includegraphics[width=.8\textwidth]{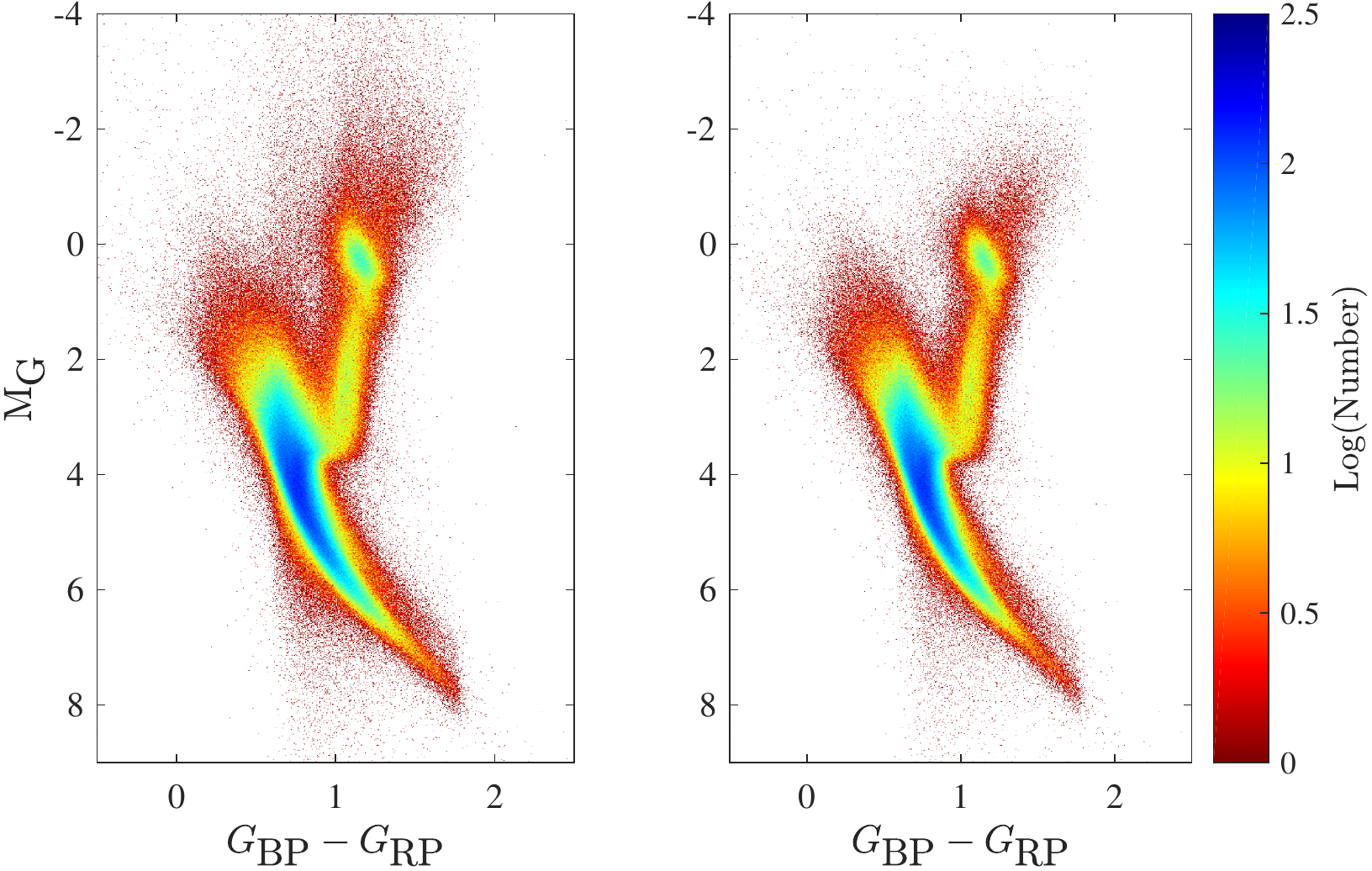}
  \caption{M$_G$ vs $G_{\textrm{BP}} - G_{\textrm{RP}}$ distribution for the full UV star sample (left panel) and
          for the stars with the relative uncertainty of $L_{\rm{UV}}$ less than 40\% (right panel). The color bar stands for the density in log scale.}
  \label{HRGaia} 
\end{figure}

Galactic extinction plays a much larger role for the FUV/NUV than the other bands \citep{Cardelli89}. In order to correct the extinction,
we use the 3D dust map from \citet{Green15}, which gives $E(B-V)$ as a function of the distance and the position. In conjunction with Gaia
parallax derived distances, we estimate the reddening in the line of sight for each star in our sample. We adopt the extinction coefficients
from \citet{Yuan13} and \citet{Jordi10} for the FUV/NUV and Gaia bands, respectively. All the magnitudes presented hereafter are extinction-corrected.
The HR diagram of our sample (Fig. \ref{HRGaia}, left panels) contains 1,474,479 UV stars with valid distances and extinctions,
in which the magnitudes and colors are extracted from the Gaia DR2. 

\begin{figure}
   \centering
    \includegraphics[width=.9\textwidth]{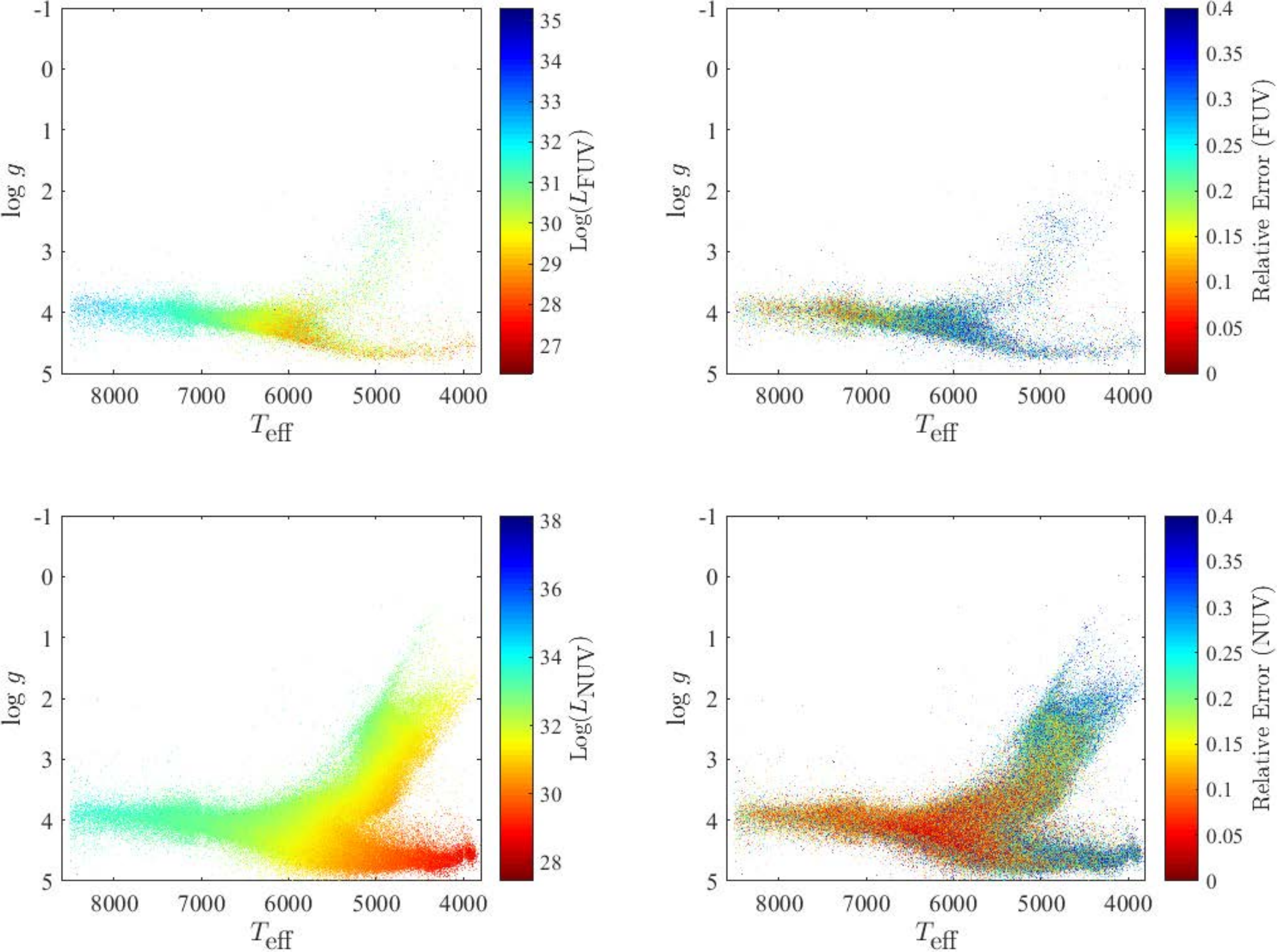}
  \caption{HR-like diagrams.The color bars stand for the luminosities in the FUV and NUV in the left two panels
           and the relative uncertainties of the luminosites in the right two panels. }
  \label{HR_like} 
\end{figure}

Using the trigonometric parallax from the Gaia DR2, we calculate the luminosity in the FUV/NUV.
We select the stars with the relative uncertainty of $L_{\rm{UV}}$ less than 40\%, and find that their $\sigma_{\pi}$/$\pi < $ 20\%.
In this case, the LKE in our sample does not significantly influence the luminosity.
There are 85,444 and 1,271,863 stars satisfied the criteria in the FUV and NUV, respectively (Fig. \ref{HRGaia}, right panel).
The number of stars with M$_{G} < -2$ are less than
those in the full sample; this may be due to the LKE in which the absolute magnitudes are over estimated.
We present the HR-like diagram of these stars in the Fig. \ref{HR_like}. The stars become luminous with increasing
effective temperatures, which is similar to the results of \citet{Shkolnik13} and \citet{Bai18}.

\section{Result}
\label{sec:res}
\subsection{UV luminous stars}
\begin{figure}
   \centering
    \includegraphics[width=.9\textwidth]{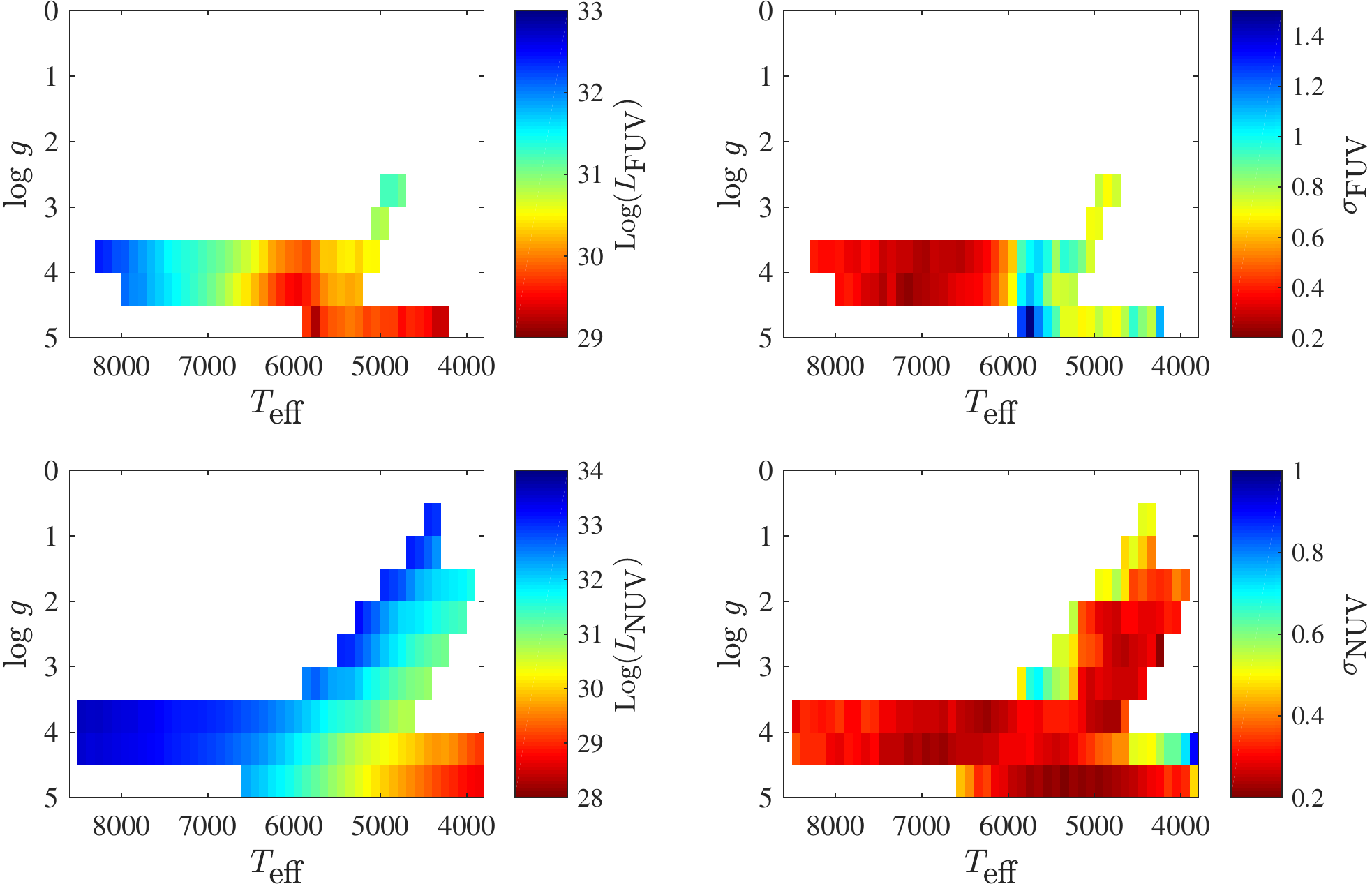}
  \caption{Distributions of the fit results in the FUV (left panel) and the NUV (right panel).
           The color bars stand for the $\mu$ and the $\sigma$ for the Gaussian fit in the bins }
  \label{HRbin_like}
\end{figure}

This luminosity catalog provides a wealth of information for the study of stellar UV emission, and enables us to investigate the
UV reference frame. We bin the stars in the log $g$ vs $T_\textrm{eff}$ diagram with the steps of 0.5 in log $g$ and 100 K in $T_\textrm{eff}$.
The distribution of the luminosity in each bin is fitted with a Gaussian function, if the number of the stars  in the bin is more than 200.
We then select the stars fall outside the normal ranges (5$\sigma$) of the Gaussian function.
Here the $\sigma$ stands for the standard deviation of the Gaussian fit.
The fitting results are shown in Fig. \ref{HRbin_like}, Table \ref{Tab:fFUV} and \ref{Tab:fNUV}.
The selection yields 275 luminous and 17 quiet stars in the FUV, and 1,774 luminous and 943 quiet stars in the NUV.
The luminosity for these stars is listed in Table \ref{Tab:sFUV} and \ref{Tab:sNUV}.

\begin{figure}
   \centering
    \includegraphics[width=1.1\textwidth]{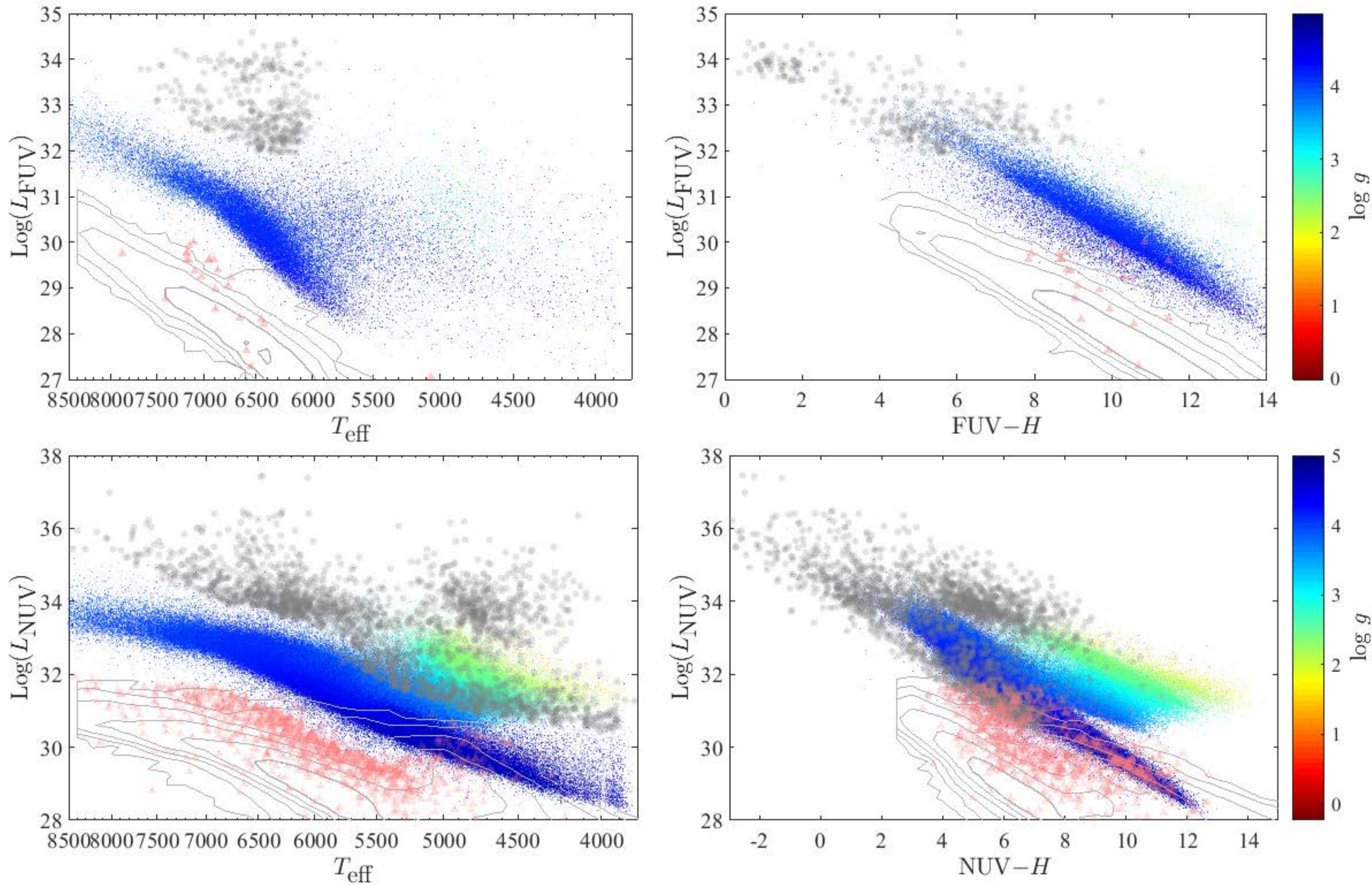}
  \caption{UV luminosity vs $T_\textrm{eff}$ (left panels) and UV $- H$ color (right panels). The
           UV luminous stars are grey points, and the UV quiet stars are pink triangles. The contours
           stand for the distributions from the BT-Cond grid \citep{Allard10} of the PHOENIX photospheric model \citep{Hauschildt99}.
           The colorbars are the values of log $g$.  }
  \label{HR}
\end{figure}

We plot the luminosity vs $T_\textrm{eff}$ in the left panels of Fig. \ref{HR}. The UV luminous and quiet stars are located along the
main sequence, when $T_\textrm{eff} \gtrsim$ 5500 K. For the stars with $T_\textrm{eff} \lesssim$ 5500 K, since they are no longer dominated by dwarfs,
the distributions for the UV luminous and quiet stars aren't located along the main sequence. We find that the UV quiet stars are
consistent with the theocratical model (the contours in Fig. \ref{HR}). These stars don't shown obvious UV excesses
above the emission predicted by the model. The luminosity vs F(N)UV $- H$ is shown in the right panels in Fig. \ref{HR}. The UV luminous
and quiet stars don't distribute along the main sequence, because the UV $-$ IR colors of the early type stars depend on the $T_\textrm{eff}$ \citep{Bai18}.
Again, the color of the UV quiet stars are consistent with the theoretical colors from the model.

\begin{table}
\center
\caption[]{Fit Result in the FUV.\label{Tab:fFUV}}
\begin{threeparttable}
 \begin{tabular}{rccrcrcc}
  \hline
Ind&$T_\textrm{eff}$&log $g$&$\sigma$&$\mu$&Num&HighNum&LowNum \\
(1)&(2)&(3)&(4)&(5)&(6)&(7)&(8)\\
  \hline
  1&4250&4.75&1.11$\pm$0.11&29.30$\pm$0.10&   205& 0& 0\\
  2&4350&4.75&0.77$\pm$0.06&29.29$\pm$0.06&   243& 0& 0\\
  3&4450&4.75&0.84$\pm$0.09&29.51$\pm$0.08&   245& 0& 0\\
  4&4550&4.75&0.95$\pm$0.06&29.59$\pm$0.06&   243& 0& 0\\
  5&4650&4.75&0.78$\pm$0.06&29.64$\pm$0.06&   312& 0& 0\\
  6&4750&2.75&0.75$\pm$0.07&31.07$\pm$0.07&   260& 0& 0\\
  7&4750&4.75&0.69$\pm$0.08&29.52$\pm$0.08&   337& 0& 0\\
  8&4850&2.75&0.69$\pm$0.06&31.23$\pm$0.06&   253& 0& 0\\
  9&4850&4.75&0.71$\pm$0.03&29.73$\pm$0.03&   372& 0& 0\\
 10&4950&2.75&0.76$\pm$0.04&31.23$\pm$0.04&   200& 0& 0\\
 11&4950&3.25&0.71$\pm$0.08&30.79$\pm$0.08&   216& 0& 0\\
 12&4950&4.75&0.77$\pm$0.04&29.74$\pm$0.04&   440& 0& 0\\
 13&5050&3.25&0.70$\pm$0.03&30.86$\pm$0.03&   226& 0& 1\\
 14&5050&3.75&0.74$\pm$0.04&30.48$\pm$0.04&   212& 0& 0\\
  \hline
\end{tabular}
\begin{tablenotes}
        \footnotesize
        \item
        (1). The index of the bin.
        (2). The median $T_\textrm{eff}$ in the unit of K.
        (3). The median log $g$.
        (4). The standard deviation of the Gaussian fit in erg s$^{-1}$.
        (5). The center of the distribution from the Gaussian fit in erg s$^{-1}$.
        (6). The number of stars in the bin.
        (7). The number of stars with the lower limit of luminosity 5$\sigma$ higher than the center of the Gaussian fit.
        (8). The number of stars with the higher limit of luminosity 5$\sigma$ lower than the center of the Gaussian fit. \\
        (This table is available in its entirety in electric form.)
\end{tablenotes}
\end{threeparttable}
\end{table}

\begin{table}
\center
\caption[]{Fit Result in the NUV.}\label{Tab:fNUV}
\begin{threeparttable}
 \begin{tabular}{rccrcrcc}
  \hline
Ind&$T_\textrm{eff}$&log $g$&$\sigma$&$\mu$&Num&HighNum&LowNum \\
  \hline
  1&3850&4.25&0.90$\pm$0.10&29.07$\pm$0.09&   247& 0& 0\\
  2&3850&4.75&0.47$\pm$0.08&28.69$\pm$0.07&   767& 2& 0\\
  3&3950&1.75&0.37$\pm$0.03&31.63$\pm$0.03&   397& 0& 0\\
  4&3950&4.25&0.72$\pm$0.04&29.20$\pm$0.03&   782& 0& 0\\
  5&3950&4.75&0.33$\pm$0.03&28.73$\pm$0.03&  1647&26& 0\\
  6&4050&1.75&0.40$\pm$0.03&31.78$\pm$0.03&   570& 2& 0\\
  7&4050&2.25&0.31$\pm$0.02&31.35$\pm$0.02&   361& 0& 0\\
  8&4050&4.25&0.62$\pm$0.05&29.37$\pm$0.05&   529& 0& 0\\
  9&4050&4.75&0.31$\pm$0.03&28.82$\pm$0.03&  1939&28& 0\\
 10&4150&1.75&0.34$\pm$0.02&31.83$\pm$0.02&   677& 1& 0\\
 11&4150&2.25&0.32$\pm$0.02&31.42$\pm$0.02&   796& 3& 0\\
 12&4150&4.25&0.62$\pm$0.05&29.56$\pm$0.05&   394& 0& 0\\
 13&4150&4.75&0.34$\pm$0.02&28.95$\pm$0.02&  2640&18& 0\\
 14&4250&1.75&0.33$\pm$0.02&31.89$\pm$0.02&   778& 3& 1\\
  \hline
\end{tabular}
\begin{tablenotes}
        \footnotesize
        \item
        The same to Table \ref{Tab:fFUV} but in the NUV \\
        (This table is available in its entirety in electric form.)
\end{tablenotes}
\end{threeparttable}
\end{table}

\begin{table}
\center
\caption[]{The FUV luminous and quiet stars.}\label{Tab:sFUV}
\begin{threeparttable}
 \begin{tabular}{rrcccrcc}
  \hline
Ind&Obsid&log($L_\textrm{FUV}$)&log(e$L_\textrm{FUV}$)&Flag&Type&Catalog&WDMS \\
(1)&(2)&(3)&(4)&(5)&(6)&(7)&(8)\\
  \hline
 13&  55109124&27.07&26.65&0&Star& &    \\
 41& 128907057&32.86&32.20&1&Star& &    \\
 42& 133409048&33.28&32.48&1&RR&D&    \\
 42& 231312031&32.76&32.26&1&RR&D&    \\
 43& 260814117&34.40&33.48&1&Hot subdwarf& &    \\
 44&  31410196&32.65&32.13&1&RR&D&    \\
 44&  75313026&32.53&32.07&1&RR& &    \\
 44& 145411088&32.50&31.92&1&RR&D&    \\
 44& 148715114&32.65&32.12&1&   &D&    \\
 44& 154611186&32.72&32.17&1&RR& &    \\
 44& 189215005&33.45&32.96&1&RR&D&    \\
 44& 218501180&32.36&30.85&1&Star& &1\\
  \hline
\end{tabular}
\begin{tablenotes}
        \footnotesize
        \item
        (1). The index of the bin that the stars are located in.
        (2). The LAMOST obsid.
        (3). The luminosity in erg s$^{-1}$.
        (4). The uncertainty of the luminosity in erg s$^{-1}$.
        (5). The stars with the lower limit of luminosity higher than the center the Gaussian fit by 5$\sigma$, Flag $=$ 1, and
             the stars with the higher limit of luminosity lower than the center by 5$\sigma$, Flag $=$ 0.
        (6). The types from the Simbad archive data.
        (7). The catalog flag. 'D'-- \citet{Drake13}, 'S'-- \citet{Sesar10}, and 'A'-- \citet{Abbas14}.
        (8). The flag of the WDMS candidates. \\
        (This table is available in its entirety in electric form.)
\end{tablenotes}
\end{threeparttable}
\end{table}

\begin{table}
\center
\caption[]{The NUV luminous and quiet stars.}\label{Tab:sNUV}
\begin{threeparttable}
 \begin{tabular}{rrcccrcc}
  \hline
Ind&Obsid&log($L_\textrm{NUV}$)&log(e$L_\textrm{NUV}$)&Flag&Type&Catalog&WDMS \\
  \hline
  9& 241109013&30.96&29.17&1&Binary& &1\\
  9& 335113059&30.74&29.99&1&Star  & &1\\
 19& 318815093&27.92&27.34&0&High proper-motion Star& &    \\
 30& 267316208&33.62&33.01&1&Eclipsing binary of Algol type& &    \\
 30& 286203005&32.80&32.13&1&   & &    \\
 30& 269109099&29.77&29.21&0&   & &    \\
 32& 321511220&32.50&31.64&1&White Dwarf& &1\\
 40& 210907122&30.86&29.42&1&X-ray source& &1\\
111& 217614072&34.19&33.76&1&RR&D&    \\
121& 151015067&33.65&32.89&1&RR&DA&  \\
  \hline
\end{tabular}
\begin{tablenotes}
        \footnotesize
        \item
        The same to Table \ref{Tab:sFUV} but in the NUV \\
        (This table is available in its entirety in electric form.)
\end{tablenotes}
\end{threeparttable}
\end{table}

\begin{figure}
   \centering
    \includegraphics[width=0.6\textwidth]{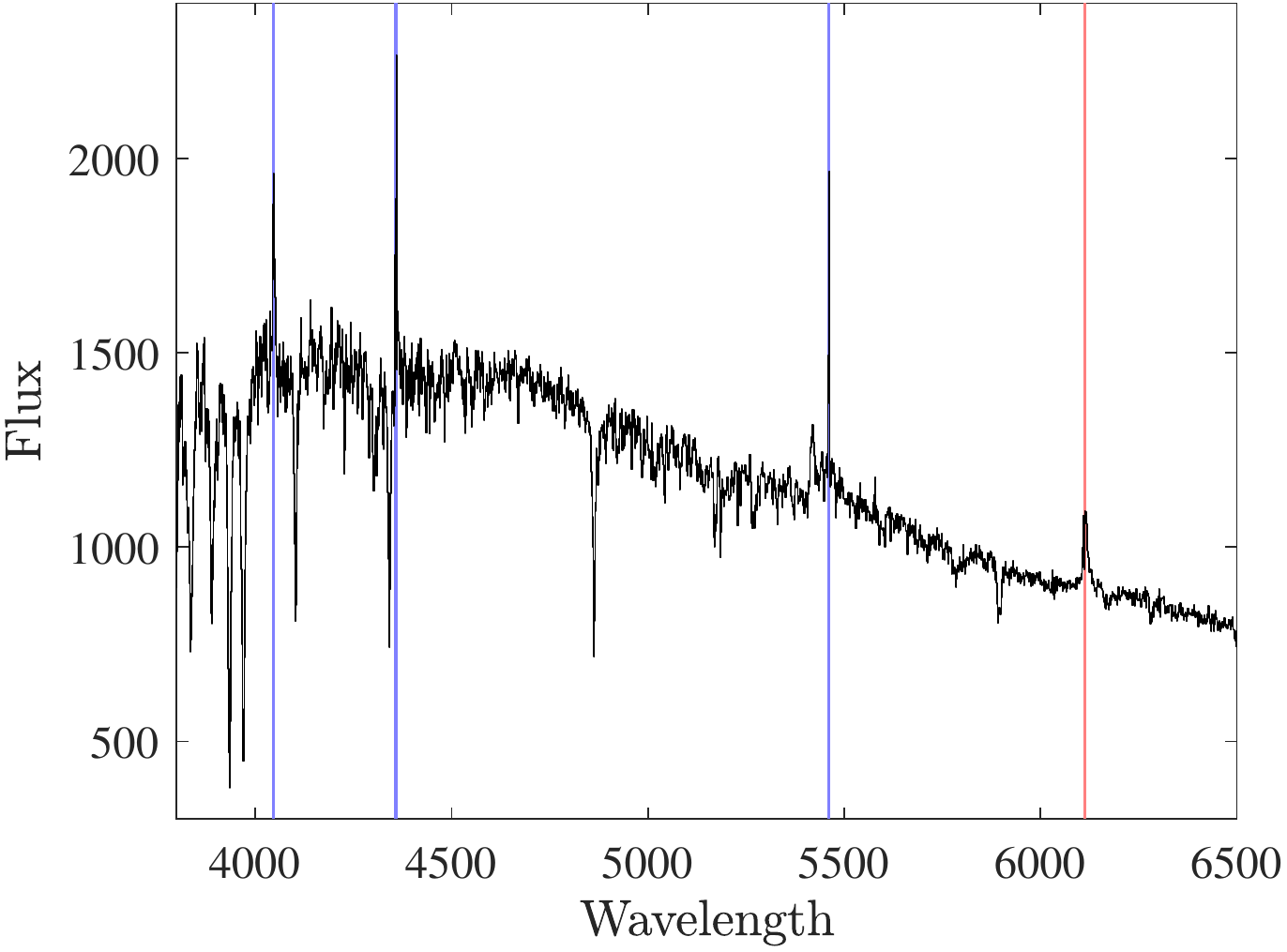}
  \caption{LAMOST spectrum of a UV luminous star (Obsid $=$ 167801105). The red lines mark the Hg I emission lines,
           and the blue line is the Fe II emission line.  }
  \label{exp}
\end{figure}

\subsection{What are these stars? }
RR Lyrae (RR Lyr) stars are pulsating periodic horizontal branch variables with great variation in the UV with a range of
2¨C5 mag \citep{Wheatley12}, making them more likely to be strong UV emitters.
We cross-match the UV luminous stars with the RR Lyr catalogs in \citet{Sesar10}, \citet{Drake13}, \citet{Abbas14}
and the Simbad archive data with a match radius of {2\arcsec}.
There are 46 and 69 RR Lyr stars that are luminous in the FUV and NUV, respectively. The chances to match a RR Lyr star are $\sim$ 17\% and $\sim$ 4\%
in the FUV and NUV of our sample, higher than that in the UV star catalog, $\lesssim$ 0.1\% \citep{Bai18}, indicating that
the RR Lyr stars are likely to emit excessive UV photons.

The binaries composed of non-degenerate stars and white dwarfs are probably unresolved by LAMOST spectra, and
these binaries may become more luminous in the UV than that expected for the secondary stars.
\citet{Bai18} presented a sample of the potential white dwarf main-sequence binaries based on their density distribution in the
FUV $-$ NUV vs $W$1 $-$ $W$2, which are extracted from $GALEX$ and $WISE$ catalogs.
We cross-match the UV luminous stars with this sample, and find 123 and 213 stars in the FUV and NUV corresponding to
$\sim$ 45\% and $\sim$ 12\% of our sample. If these stars are white dwarf binaries, their UV luminosities are
dominated by the accretion disks or the hot atmospheres of the white dwarfs.

The FUV $-$ NUV vs $W$1 $-$ $W$2 diagram are powerful to select white dwarf plus M dwarf binaries, since their emissions are mainly in the
UV and IR bands, respectively. For binaries with secondary stars with earlier than M, the IR color is not sensitive enough to distinguish
white dwarf binaries from single stars.
We present an example of a UV luminous star in Fig. \ref{exp}. It is a spectrally identified F star with $T_\textrm{eff} =$ 6298 K, and
log $g =$ 4.05. Its FUV luminosity exceeds the center of the reference frame by 5.9$\sigma$.
The three emission lines marked in blue are the nigh sky emission line from mercury.  The red emission line is Fe II $\lambda$6113
originated from hot thin gas, which has been detected in the nova spectra, e.g., Cyg 2014 \footnote{http://www.astrosurf.com/aras/Aras\_DataBase/DataBase.htm}
and V445 Puppis \citep{Iijima08}. The star probably harbors in a binary system around a spectrally unresolved white dwarf that raises the
FUV emission and the Fe II line. This implies that there are some potential white dwarf $-$ G dwarf binaries and white dwarf $-$ F dwarf binaries
in our sample, which require additional optical colors to isolated them.

There may be also some active stars or active non-degenerate binaries in our sample. They are probably fast rotating stars with
strong stellar activity \citep{Reiners12}, or they may have the interconnecting field lines with the companion \citep{Simon80}.
They also could be very young stars or hot subdwarfs with UV excessive emission from hot atmospheres.
The neutron star and the black hole binaries aren't ruled out, but we need multi-band information from radio to X ray to identify them.

\section{Summary}
\label{sec:sum}
We present a catalog of 275 and 1,774 luminous stars in the FUV and NUV, which exceed the center of the
local reference frame with at least 5$\sigma$. The reference frame is built with over one million LAMOST UV stars in log $g$ vs $T_\textrm{eff}$
diagram. We correct the extinction for the UV emission using the 3D dusty map with the distances from the Gaia DR2.
These UV luminous stars are selected with the relative uncertainties of the luminosity less
than 40\% and the trigonometric parallax less than 20\%, therefore they don't suffer significant LKE. We cross-match our sample with the catalogs
of RR Lyr stars and possible white dwarf main-sequence binaries, and they in all
compose $\sim$ 62\% and $\sim$ 16\% of our sample in the FUV and NUV, respectively.

This catalog provides an ideal sample to study stellar activity, compact main-sequence binaries and variable stars.
These objects probably have optical spectra similar to normal stars, but have abnormal emission in the UV.
We are going to use the most recent data release of the LAMOST to enlarge the sample, and study them in detail to
further shed light on the nature of these UV luminous stars.

\begin{acknowledgements}
This work was supported by the National Program on Key Research and Development
Project (Grant No. 2016YFA0400804) and
the National Natural Science Foundation of China (NSFC)
through grants NSFC-11603038/11333004/11425313/11403056.
Some of the data presented in this paper were obtained from the Mikulski Archive for
Space Telescopes (MAST). STScI is operated by the Association of Universities for Research
in Astronomy, Inc., under NASA contract NAS5-26555. Support for MAST for non-HST data is provided
by the NASA Office of Space Science via grant NNX09AF08G and by other grants and contracts.

The Guoshoujing Telescope (the Large Sky Area Multi-Object
Fiber Spectroscopic Telescope, LAMOST) is a National Major
Scientific Project which is built by the Chinese Academy of
Sciences, funded by the National Development and Reform Commission,
and operated and managed by the National Astronomical Observatories,
Chinese Academy of Sciences.

This work has made use of data from the European Space Agency (ESA) mission
{\it Gaia} (\url{https://www.cosmos.esa.int/gaia}), processed by the {\it Gaia}
Data Processing and Analysis Consortium (DPAC,
\url{https://www.cosmos.esa.int/web/gaia/dpac/consortium}). Funding for the DPAC
has been provided by national institutions, in particular the institutions
participating in the {\it Gaia} Multilateral Agreement.
\end{acknowledgements}

\label{lastpage}

\end{document}